\documentclass[twocolumn,showpacs,preprintnumbers,amsmath,amssymb,superscriptaddress,graphicx,graphics,pstricks]{revtex4}
\usepackage{amsmath}
\usepackage{graphicx}
\usepackage[latin1]{inputenc}
\usepackage{pstricks}
\usepackage{pst-node}
\usepackage{pst-coil}
\usepackage{pst-grad}
\usepackage{multido}
\usepackage{dcolumn}
\usepackage{bm}

\begin{document}

\title{Finite size effects and non-additivity in the van der Waals interaction}

\author{Reinaldo de Melo e Souza}
\affiliation{Instituto Federal do Rio de Janeiro, Rua Lucio Tavares 1045, Nil\'opolis, Rio de Janeiro, RJ,  26530-060, Brasil}
\author{W.J.M. Kort-Kamp}
\author{C. Sigaud}
\author{C. Farina}
\affiliation{Instituto de Fisica, UFRJ, CP 68528, Rio de Janeiro,
RJ, 21941-972, Brasil}

\begin{abstract}
We obtain analytically the exact non-retarded dispersive interaction energy between an atom and a perfectly conducting disc. We consider the atom in the symmetry axis of the disc and assume the atom is predominantly polarizable in the direction of this axis. For this situation we discuss the finite size effects on the corresponding interaction energy. We follow the recent procedure introduced by Eberlein and Zietal together with the old and powerful Sommerfeld's image method for non-trivial geometries. For the sake of clarity we present a detailed discussion of Sommerfeld's image method.
Comparing our results form the atom-disc system with those recently obtained for an atom near a conducting plane with a circular aperture, we discus the non-additivity of the van der Waals interactions involving an atom and two complementary surfaces. We show that there is a given ratio $z/a$ between the distance $z$ from the atom to the center of the disc (aperture) and the radius of the disc $a$ (aperture) for which non-additivity effects vanish. Qualitative arguments suggest that this quite unexpected result will occur not only for a circular hole, but for anyother symmetric hole.
 \end{abstract}

\maketitle
%
%
%
%
\section{Introduction}

Motivated by the growing interest in the computation of dispersive forces between a polarizable atom, as well as small conducting objects like a needle or a small cylinder, and non-trivial conducting surfaces
\cite{McCauley2011-B,McCauley2011,Levin2010,Milton2011,Maghrebi2011,Eberlein2011,Eberlein2009,Eberlein2007}, to mention just a few recent works on the subject, we present the analytic solution for the non-retarded interaction energy between an atom and a perfectly conducting disc. Particularly, our interest in the atom-disc system  relies on the fact that this configuration is somehow complementary to that of an atom  interacting with an infinite plane with a circular hole, recently discussed in the literature \cite{Levin2010,Eberlein2011,McCauley2011}. It is worth mentioning that  Casimir forces between complementary conducting surfaces have been discussed recently with the aid of the scattering formula in connection with Babinet principle \cite{Maghrebi2011b}. In this letter, our purposes are essentially to analyze the finite size effects in the atom-disc system and to discuss some aspects of non-additivity effects in the van der Waals interaction involving the two complementary geometries mentioned above. For our purposes, suffice it to consider the atom in the symmetry axis of the disc and assume the atom predominantly polarizable in the direction of this axis. Curious as it may seem, we show that there is a ratio $z/a$ between the distance $z$ from the atom to the center of the disc (hole) and the radius of the disc $a$ (hole) for which the non-additivity effects vanish. Further, based on qualitative arguments, we conjecture that this unexpected vanishing of the non-additivity effects will occur for complementary surfaces independently of the form of the hole.
This paper is organized as follows: in section II we review Eberlein-Zietal method in the extremely simple case of an atom near an infinite perfectly conducting plane just to emphasize the important role played by the image method, a crucial step in the approach we shall  employ later on in more involved situations. In section III we use Sommerfeld's image method to work out the atom-disc system. In this section, we also show how to obtain the van der Waals interaction between an atom and a perfectly conducting surface with a circular hole, exploring the fact that these two surfaces are complementary to one another. With analytical results in hand, we analyze the finite size effects in the atom-disc system.
In section 4, we use the obtained results for those complementary surfaces to discuss
non-additivity effects in the van der Waals interaction. Last section is left for conclusions and final remarks.

%
%
\section{Eberlein-Zietal method and the image method}
%
%

Consider a polarizable atom at position $\mathbf{r}_0$ in the presence of a grounded perfectly conducting surface.
In a recent publication, C. Eberlein and R. Zietal \cite{Eberlein2007} have shown
that the non-retarded dispersive interaction between the atom and the surface, in first order perturbation theory,
can be obtained by taking the vacuum expectation value of the following operator
\begin{equation}
 \hat{V} = \frac{1}{2\varepsilon_0}(\hat{\mathbf{d}}\cdot\nabla')(\hat{\mathbf{d}}\cdot\nabla)G_H(\mathbf{r},\mathbf{r}')
 \bigg|_{\mathbf{r}=\mathbf{r}'=\mathbf{r}_0} \, , \label{eberlein}
\end{equation}
where $\hat{\mathbf{d}}$ is the atomic dipole operator and
$G_H(\mathbf{r},\mathbf{r}')$ satisfies Laplace's equation and a boundary condition at the surface,
%
%
\begin{eqnarray}\label{GH}
	\nabla^2 G_H ({\bf r}, {\bf r}^{\,\prime})&=&0 \\
	\left[\frac{1}{4\pi|\mathbf{r}-\mathbf{r}'|} + G_H(\mathbf{r},\mathbf{r}')\right]_{\mathbf{r}\in S}&=&0 \,.\label{CCGH}
\end{eqnarray}
Equation (\ref{eberlein}) has been applied in a variety of interesting geometries \cite{Eberlein2007,Eberlein2011,Contreras-Eberlein-2009}.

Looking at the previous equations we see that $G_H ({\bf r}, {\bf r}^{\,\prime})$ is the homogeneous solution of the non-homogeneous equation
$\nabla^2 G ({\bf r}, {\bf r}^{\,\prime}) = -\delta({\bf r} - {\bf r}^{\,\prime})$ which  makes $G ({\bf r}, {\bf r}^{\,\prime})$ vanish at the surface. Hence, apart form a constant factor, $G_H ({\bf r}, {\bf r}^{\,\prime})$ is the contribution to the electrostatic potential at point ${\bf r}$ due to the surface charge density induced by a point charge located at ${\bf r}^{\,\prime}$. In other words, $G_H ({\bf r}, {\bf r}^{\,\prime})$ is the contribution of the image charges to the total potential (if the problem admits an image treatment).
Indeed, consider a charge $q$ at $\mathbf{r}'$ in the presence
of the grounded perfectly conductor $S$. The electrostatic potential of this configuration is given by
\begin{equation}
	\phi({\bf r},{\bf r}^{\,\prime}) = \frac{q}{4\pi\varepsilon_0|\mathbf{r}-\mathbf{r}'|} +
 \phi_i({\bf r},{\bf r}^{\,\prime}) \, ,
\end{equation}
where the electrostatic potential of the image charges, denoted by $\phi_i(\mathbf{r})$, is the solution of
\begin{eqnarray}\label{PhiImagem}
	\nabla^2 \phi_i({\bf r},{\bf r}^{\,\prime})&=&0 \\
	\left[\frac{q}{4\pi\varepsilon_0|\mathbf{r}-\mathbf{r}'|} +
 \phi_i({\bf r},{\bf r}^{\,\prime})\right]_{\mathbf{r}\in S}&=&0 \,.\label{CCPhiImagem}
\end{eqnarray}
Comparing equations (\ref{GH}) and (\ref{CCGH}) with equations (\ref{PhiImagem}) and (\ref{CCPhiImagem}), we immediately conclude that
\begin{equation}
G_H(\mathbf{r},\mathbf{r}') = \frac{\varepsilon_0\phi_i({\bf r},{\bf r}^{\,\prime})}{q} \, . \label{ghim}
\end{equation}
In other words, to find the quantum dispersive force between a polarizable atom and an arbitrary grounded conducting surface in the non-retarded regime, we need only to solve an electrostatic image problem. Then, we just apply equation (\ref{eberlein}) recalling that $G_H(\mathbf{r},\mathbf{r}')$ is the image contribuition as explained above, see equation (\ref{ghim}).

We shall explore this remark in order to obtain $G_H(\mathbf{r},\mathbf{r}')$ for the proposed atom-disc system, with which
 by employing equation (\ref{eberlein}) we shall obtain the desired  van der Waals interaction energy for this system.
To illustrate clearly our procedure, let us  begin with the simplest system composed by an atom and an infinite grounded conducting plane.
The electrostatic potencial of the image of a charge $q$ at position $\mathbf{r'}=(x',y',z')$ in front of a plane, which
we set at $z=0$ for convenience, is given by
\begin{equation}\label{phii}
	\phi_i({\bf r},{\bf r}^{\,\prime}) = - \frac{q}{4\pi\varepsilon_0|\mathbf{r}-\mathbf{r}_i|} \, ,
\end{equation}
where ${\bf r}_i := {\bf r}^{\,\prime} - 2z^{\,\prime}\,\hat{\bf z}$  is the position of the image charge.
>From equations (\ref{ghim}) and (\ref{phii}) we obtain
\begin{equation}\label{GHExplicito}
G_H(\mathbf{r},\mathbf{r}') = - \frac{1}{4\pi\sqrt{(x - x^{\,\prime})^2 + (y - y^{\,\prime})^2
+ (z + z^{\,\prime})^2}}
%
\end{equation}
Hence, inserting (\ref{GHExplicito}) into (\ref{eberlein}) and taking its quantum expectation value,
 we immediately reobtain the well known result (cf. \cite{Lennard1932}, \cite{Cohen1979}) for the dispersive
interaction $E_0$ of the atom with the infinite conducting plane,
\begin{equation}
	E_{0} = -\frac{\langle d_x^2\rangle+\langle d_y^2\rangle+2\langle d_z^2\rangle}{64\pi\epsilon_0|z_0|^3} \, ,
\end{equation}
where $z_0$ is the $z$-component of the atomic position.

At first glance, the problem of a point charge in the presence of a finite conducting disc does not seem to offer an approach based
 on the image method, since there are no points in the space where we may put an image charge (remember that the image charges can not be
 put in the physical region). However, a clever procedure developed by Sommerfeld allows the generalization of the image method for dealing with such non-trivial geometries, as for example, a point charge in front of a conducting disc and a point charge in front of an infinite conducting plane with a circular hole, among others. In the next section we will show in detail how to use  Sommerfed's image method and Eberlein-Zietal in the above mentioned non-trivial geometries.

%
%
\section{Atom-disc non-retarded interaction}

Recently, an exact analytical expression for the non-retarded interaction energy between an atom and an infinitely conducting plane with a circular aperture was presented by Eberlein and Zietal \cite{Eberlein2011}. This system is quite interesting and exhibits peculiar features. Apart from involving a non-trivial geometry, for the case where the atom is in the axis of the circular hole and close enough to its center the non-retarded dispersive force on the atom will be repulsive, provided the atom is predominantly polarizable along the axis (this result had already been pointed out based on numerical computation by Levin {\it et al} \cite{Levin2010}). These papers motivated us to consider the atom in front of the complementary surface, namely, the atom-disc system. Hence, we shall consider here a polarizable atom in the axis of a perfectly conducting disc of radius $a$ and fixed at a distant $z$ above its center. By assumption, the atom-disc separation is much smaller than the dominant transition wavelength of the atom, since this is the regime of greatest practical interest as emphasized in \cite{Eberlein2011}. We know that the hypothesis of a perfectly conducting disc should be modified for short distances to describe real metals, but we shall maintain such an idealized condition otherwise we will not be able to obtain an exact analytical solution for this problem. Besides, the main features of the dispersive interaction will not change substantially so that the advantages of getting an exact analytical result will be worthwhile.
To derive our analytical expression for the non-retarded interaction energy in the atom-disc system, we combined the powerful image method devised by Sommerfeld in 1896 \cite{Sommerfeld1896} with the Eberlein-Zietal method above mentioned.

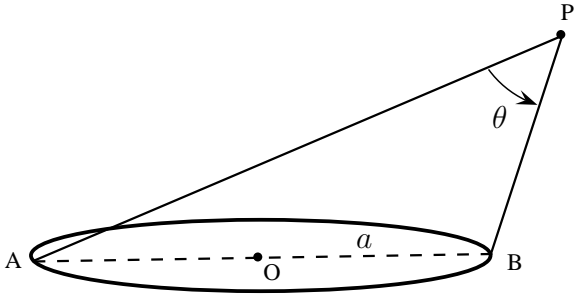
\begin{figure}[!h]
\begin{center}
\newpsobject{showgrid}{psgrid}{subgriddiv=1,griddots=10,gridlabels=6pt}
\begin{pspicture}(1,-1.6)(7,1.7)
 \psset{arrowsize=0.16 2}
\psset{unit=1}
\psellipse[linewidth=0.05,dimen=outer](3.42,-1.34)(3.08,0.5)
\psline[linewidth=0.03cm](0.38,-1.42)(7.42,1.58)
\psline[linewidth=0.03cm](6.46,-1.38)(7.44,1.62)
\psdots[dotsize=0.12](3.38,-1.36)
\usefont{T1}{ptm}{m}{n}
\rput(3.57,-1.555){O}
\usefont{T1}{ptm}{m}{n}
\psdots[dotsize=0.12](7.41,1.6)
\rput(7.5,1.9){P}
\usefont{T1}{ptm}{m}{n}
\rput(0.13,-1.395){A}
\usefont{T1}{ptm}{m}{n}
\rput(6.8,-1.435){B}
\usefont{T1}{ptm}{m}{n}
\psarc[linewidth=0.02]{->}(7.5,1.9){1.3}{216}{252}
\rput(4.8,-1.15){\large$a$}

\rput(6.6,0.5){\large$\theta$}
\psline[linewidth=0.03cm,linestyle=dashed,dash=0.16cm 0.16cm](0.4,-1.42)(6.5,-1.32)
\end{pspicture}
\end{center}
\caption{The figure shows the line segments $\overline{PA}$ and $\overline{PB}$, used in the definition of the peripolar
coordinate $\rho = \ln{\overline{PA}/\overline{PB}}$ and the peripolar coordinate $\theta$ defined for a disc of radius $a = \overline{AB}/2$.}
\label{FigPeripolar}
\end{figure}

The appropriate coordinates to analyze the atom-disc system (and an atom and an infinite plate with a circular hole as well) are the so called peripolar coordinates, introduced by C. Neumann \cite{Neumann}, \cite{Hobson1899}, defined as follows: consider the plane that contains
 a generic space point $P$  and the symmetry axis of the disc. Let $A$ and $B$ be the points at which such a plane intersects the circumference of the disc. By convention, $A$ and $B$ are chosen in a manner such that, to make the line $AP$ coincide with the line $BP$, we must
turn it in the counterclockwise sense. The peripolars coordinates of $P$ are the angle $\theta=\hat{APB}$, the number $\rho=\ln\frac{\overline{PA}}{\overline{PB}}$ and  the usual azimuthal angle $\varphi$ (the same as in cylindrical coordinates), see Figure \ref{FigPeripolar}.

The convenience of this system is evident once we write the equation of the conducting surface. It is easy
to see that $\theta=\pm\pi$ are the equations for the points belonging to the disc (see Figure \ref{FigPeripolar}).
As in the case of the atom-plane, the first thing we should do is to find the classical electrostatic potential created by a point
charge in the presence of a grounded conducting disc. This is possible following Sommerfeld's procedure, which
consists in making a copy of the ordinary space, where we may put the necessary image charges (as we will see, in the present case, we shall need
 only one image charge). With this goal, we consider the disc as a branch surface. For $-\pi<\theta<\pi$ we are in the ordinary
space, while points with $\pi<\theta<3\pi$ are in the imaginary space (auxiliary space). With this construction
we are brought back to the same point after a rotation of $4\pi$ but not after a rotation of $2\pi$. Crossing the conductor, takes us to a new space.
Now, our problem consists in a point charge $q$ at the point $\mathbf{r}' = (\rho',\theta',\varphi')$, with
$-\pi<\theta<\pi$, in the presence of a conductor in $\theta=\pm\pi$. The details of the method
are to be found in \cite{Hobson1899}, \cite{Davis1971} and \cite{Alzofon2004}. Since this method
is not well-known today and once we followed an approach that is not one of the three above but rather a mixture of them,
we find instructive to expose it here.

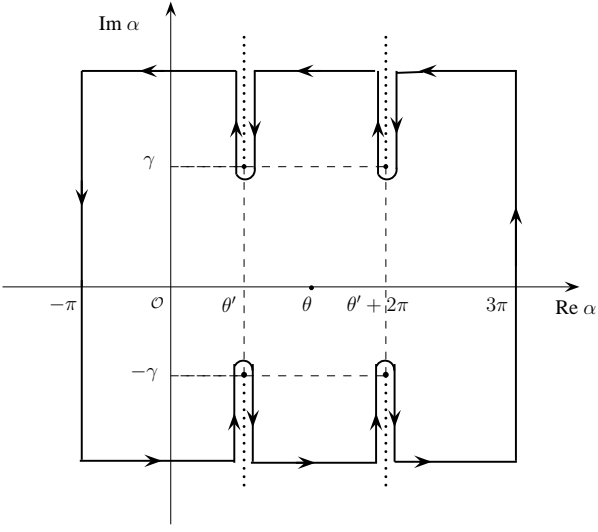
\begin{figure}
\begin{center}
\newpsobject{showgrid}{psgrid}{subgriddiv=1,griddots=10,gridlabels=6pt}

\scalebox{0.65} 
{
\begin{pspicture}(1,-5.0)(13.64,5.7)

 \psset{arrowsize=0.18 2}

\psline[linewidth=0.02cm]{->}(4.44,-5.5)(4.44,5.2)
\psline[linewidth=0.02cm]{->}(1.0,-0.64)(12.8,-0.64)
\psline[linewidth=0.04cm](2.62,-0.64)(2.62,1.54)
\psline[linewidth=0.04cm]{<-}(2.62,1.0)(2.62,3.78)
\psline[linewidth=0.04cm](2.62,3.78)(4.3,3.78)
\psline[linewidth=0.04cm]{<-}(3.8,3.78)(5.78,3.78)
\psline[linewidth=0.04cm](5.78,3.78)(5.78,2.68)
\psline[linewidth=0.04cm]{<-}(5.78,2.78)(5.78,1.66)
\psline[linewidth=0.04cm](6.158,1.78)(6.158,2.6)
\psline[linewidth=0.04cm]{<-}(6.158,2.4)(6.158,3.78)
\psline[linewidth=0.06cm,linestyle=dotted,dash=0.16cm
0.16cm](5.94,4.5)(5.94,1.82)
\psline[linewidth=0.02cm,linestyle=dashed,dash=0.16cm
0.16cm](5.94,1.82)(5.94,-2.44)
\psline[linewidth=0.06cm,linestyle=dotted,dash=0.16cm
0.16cm](5.94,-2.44)(5.94,-4.72)
\rput{-180.0}(11.92,3.52)
{\psarc[linewidth=0.04](5.96,1.76){0.2}{30.068583}{191.30994}}
\psdots[dotsize=0.1](5.94,1.82)
\psline[linewidth=0.02cm,linestyle=dashed,dash=0.16cm
0.16cm](4.44,1.82)(5.9,1.82)
\psline[linewidth=0.04cm](5.74,-2.22)(5.74,-3.32)
\psline[linewidth=0.04cm]{<-}(5.74,-3.22)(5.74,-4.24)
\psline[linewidth=0.04cm](6.118,-4.22)(6.118,-3.4)
\psline[linewidth=0.04cm]{<-}(6.118,-3.52)(6.118,-2.22)
\rput{-35.0}(2.4117162,2.974636)
{\psarc[linewidth=0.04](5.9230285,-2.3371754){0.18766797}{30.068583}{191.30994}}
\psdots[dotsize=0.12](5.94,-2.44)
\psline[linewidth=0.02cm,linestyle=dashed,dash=0.16cm
0.16cm](4.44,-2.46)(5.9,-2.46)
\psline[linewidth=0.04cm]{<-}(7,3.78)(8.62,3.78)
\psline[linewidth=0.04cm]{-}(6.14,3.78)(7.5,3.78)

\psline[linewidth=0.04cm]{<-}(9.52,3.78)(11.5,3.78)
\psline[linewidth=0.04cm](11.5,3.8)(11.5,-0.22)
\psline[linewidth=0.04cm](2.62,-4.18)(2.62,-0.5)
\usefont{T1}{ptm}{m}{n}
\rput(12.73,-1.055){\large Re $\alpha$}
\usefont{T1}{ptm}{m}{n}
\rput(3.39,4.745){\large Im $\alpha$}
\usefont{T1}{ptm}{m}{n}
\rput(4.17,-0.975){${\cal O}$}
\usefont{T1}{ptm}{m}{n}
\rput(11.12,-0.995){\large$3\pi$}
\usefont{T1}{ptm}{m}{n}
\rput(5.64,-0.995){\large$\theta'$}
\usefont{T1}{ptm}{m}{n}
\rput(4,1.9){\large$\gamma$}
\usefont{T1}{ptm}{m}{n}
\rput(3.9,-2.4){\large$-\gamma$}
\psline[linewidth=0.04cm]{->}(6.08,-4.24)(7.4,-4.24)
\psline[linewidth=0.04cm]{-}(7,-4.24)(8.66,-4.24)

\psline[linewidth=0.04cm](9.52,-4.22)(11.52,-4.22)
\psline[linewidth=0.04cm]{->}(2.58,-4.2)(4.26,-4.2)
\psline[linewidth=0.04cm](4.14,-4.2)(5.74,-4.2)
\usefont{T1}{ptm}{m}{n}
\rput(3.65,5.505){\color{white}Oi}
\psdots[dotsize=0.1](7.32,-0.66)
\usefont{T1}{ptm}{m}{n}
\rput(7.23,-0.995){\large$\theta$}
\psline[linewidth=0.04cm]{<-}(11.5,1)(11.5,-4.2)
\psline[linewidth=0.04cm](8.68,3.78)(8.68,2.68)
\psline[linewidth=0.04cm]{<-}(8.68,2.78)(8.68,1.66)
\psline[linewidth=0.04cm](9.058,1.78)(9.058,2.6)
\psline[linewidth=0.04cm]{<-}(9.058,2.48)(9.058,3.78)
\psline[linewidth=0.06cm,linestyle=dotted,dash=0.16cm
0.16cm](8.84,4.5)(8.84,1.82)
\psline[linewidth=0.02cm,linestyle=dashed,dash=0.16cm
0.16cm](8.84,1.82)(8.84,-2.44)
\psline[linewidth=0.06cm,linestyle=dotted,dash=0.16cm
0.16cm](8.84,-2.44)(8.84,-4.72)
\rput{-180.0}(17.72,3.52){\psarc[linewidth=0.04](8.86,1.76){0.2}{30.068583}{191.30994}}
\psdots[dotsize=0.1](8.84,1.82)
\psline[linewidth=0.02cm,linestyle=dashed,dash=0.16cm
0.16cm](4.64,1.82)(8.8,1.82)
\psline[linewidth=0.04cm](8.64,-2.22)(8.64,-3.32)
\psline[linewidth=0.04cm]{<-}(8.64,-3.22)(8.64,-4.24)
\psline[linewidth=0.04cm](9.018,-4.22)(9.018,-3.4)
\psline[linewidth=0.04cm]{<-}(9.018,-3.52)(9.018,-2.22)
\rput{-35.0}(2.9361756,4.638008){\psarc[linewidth=0.04](8.823029,-2.3371756){0.18766797}{30.068583}{191.30994}}
\psdots[dotsize=0.12](8.84,-2.44)
\psline[linewidth=0.02cm,linestyle=dashed,dash=0.16cm
0.16cm](4.64,-2.46)(8.8,-2.46)
\psline[linewidth=0.04cm](9.06,3.74)(9.6,3.76)
\psline[linewidth=0.04cm]{->}(9.02,-4.22)(9.78,-4.22)
\usefont{T1}{ptm}{m}{n}
\rput(8.68,-0.995){\large$\theta'+2\pi$}
\usefont{T1}{ptm}{m}{n}
\rput(2.24,-1.015){\large$-\pi$}
\end{pspicture}
}

\end{center}
\caption{Choice of counter $C$ used in integration of (\ref{umsobrerdisco}).}
\label{Contorno}
\end{figure}

If we are to work in the double space, the coulomb potential $1/R$ cannot be the potential of a single charge,
since it has a symmetry in changing $\theta$ by $\theta+2\pi$. Henceforth, the laplacian of $1/R$ corresponds
to two Dirac delta functions, one with singularity at $\mathbf{r}'=(\rho',\theta',\varphi')$ and another with
singularity at  $\mathbf{r}'=(\rho',\theta'+2\pi,\varphi')$. This way, the potential $1/R$ represents, in the double space, the
superposition of two point charges, one in the ordinary space and another in the imaginary auxiliary space.
In order to identify each charge contribution we shall write the
distance $R$ between two any points in peripolars coordinates. This can be easily done once we establish the
relations between these coordinates and the cylindrical coordinates, $r$ and $z$. Following \cite{Hobson1899}, we can write
\begin{eqnarray}
		z&	=&\frac{a\sin\theta}{\cosh\rho-\cos\theta}\, , \\
		r^2&=&\frac{a^2\sinh\rho}{\cosh\rho-\cos\theta} \, .
\end{eqnarray}
Then, the square of the distance between the points $(\rho',\theta',\varphi')$ and $(\rho,\theta,\varphi)$ is
\begin{eqnarray}
&&	R^2=(z-z')^2+r^2+r'^2-2rr'\cos(\varphi-\varphi')=2a^2\nonumber \\
	&\times&\frac{\cosh\rho\cosh\rho'-\sinh\rho\sinh\rho'\cos(\varphi-\varphi')-\cos(\theta-\theta')}{(\cosh\rho-\cos\theta)(\cosh\rho'-\cos\theta')} \, . \nonumber \\
\end{eqnarray}
Since $\cosh\rho\cosh\rho'-\sinh\rho\sinh\rho'\cos(\varphi-\varphi')\geq 1$, it exists a real $\gamma$ such that
\begin{equation}
\cosh\gamma=	\cosh\rho\cosh\rho'-\sinh\rho\sinh\rho'\cos(\varphi-\varphi'),
\end{equation}
which allows us to write this distance in the form
\begin{equation}
	R=a\sqrt2\frac{[\cosh\gamma-\cos(\theta-\theta')]^{1/2}}{(\cosh\rho-\cos\theta)^{1/2}(\cosh\rho'-\cos\theta')^{1/2}} \, . \label{rperi}
\end{equation}
Defining
\begin{equation}
	R_{\alpha}=a\sqrt2\frac{[\cosh\gamma-\cos(\alpha-\theta')]^{1/2}}{(\cosh\rho-\cos\theta)^{1/2}(\cosh\rho'-\cos\theta')^{1/2}} \, , \label{Ralpha}.
\end{equation}
where $\alpha$ is a complex variable, we can use Cauchy's theorem to write the coulomb potential as
\begin{equation}
	\label{umsobrerdisco}
	\frac{1}{R}=\frac{1}{2\pi i}\oint\limits_{C} \frac{f(\alpha)}{R_{\alpha}} \, d\alpha \, .
\end{equation}
Function $f$ must have a singularity at $\alpha=\theta$ and unitary residue, and the contour $C$ can not enclose
any singularity other than $\alpha=\theta$. 
One may wonder why in the definition (\ref{Ralpha}) we changed $\theta$ by $\alpha$ only in the numerator. 
It is possible to change it also in the denominator. It can be shown that this procedure does not change the final result but it allows one to deform the path of integration in a convenient way, 
as shown in Figure \ref{Contorno}. For further details, see Davis and Reitz\cite{Davis1971}.
Since our double space is 4$\pi$-periodic in the $\theta$
variable, it is convenient to choose $f$ with this same period. With this in mind, an appropriate choice
seems to be
\begin{equation}
	\label{f2}
	f(\alpha)=\frac{1}{2}\frac{i}{(1-e^{i(\theta-\alpha)/2})} \, .
\end{equation}
Our contour of integration must exclude the singularities of
$1/R_{\alpha}$. The equation (\ref{Ralpha}) reveals the singularities at
\begin{equation}
	\alpha=\theta'+2m\pi \pm  i\gamma \, \, \, , m\in Z \, .
\end{equation}
We may take the contour as sketched in Figure \ref{Contorno}
%


Now we must perform the integration (\ref{umsobrerdisco}). The calculation readily exposes the
convenience of the chosen circuit. The contributions
 of the vertical lines at Re $\alpha=-\pi$ and Re $\alpha=3\pi$ cancel out due to the symmetry
of the integrand. The horizontal lines give null contributions in the limit  Im $\alpha\rightarrow\pm\infty$.
We are left with the integrations around the singularities $\alpha=\theta'\pm i\gamma$
 and $\alpha=\theta'+2\pi\pm i\gamma$.
We call the former path $A_0$ and the latter $A_1$. Then, integral (\ref{umsobrerdisco}) reads
\begin{equation}
	\label{potespdup}
		\frac{1}{R}=\frac{1}{4\pi }\int\limits_{\, A_0} \frac{R_{\alpha}^{-1}}{1-e^{i(\theta-\alpha)/2}} d\alpha +\frac{1}{4\pi }\int\limits_{\, A_1} \frac{R_{\alpha}^{-1}}{1-e^{i(\theta-\alpha)/2}} d\alpha\, .
\end{equation}
Sommerfeld has shown, cf.\cite{Sommerfeld1896}, that the first integral

\noindent
{\it (i)} is uniquely defined, finite and
continuous at all points of the double space, except at $(\rho',\theta',\varphi')$. This means, particularly,
that it is finite at $(\rho',\theta'+2\pi,\varphi')$;

 \noindent
 {\it (ii)} has a null laplacian  at all points except at $(\rho',\theta',\varphi')$ and at the surface of the conductor;

  \noindent
  {\it (iii)} vanishes at infinity and

  \noindent
  {\it (iv)} is bivalent in the ordinary space, with a separated branch to each copy of the double space. Thus, including the constant factor
$q/4\pi\varepsilon_0$, we see that
\begin{equation}
	V(\rho,\theta,\varphi)=\frac{q}{16\pi^2\varepsilon_0}\int \limits_{\, A_0} \frac{R_{\alpha}^{-1}}{1-e^{i(\theta-\alpha)/2}} d\alpha \, ,
\end{equation}
is the potential of a single charge in the double space. This integral can be easily performed. First,
note that there are two terms to be considered. One which is around the branch point at $\alpha=\theta'+ i\gamma$ and
other which is around the branch point at $\alpha=\theta'- i\gamma$. In the first case, we have three paths, namely, two verticals, one
in which Re $\alpha=\theta'-\delta$ and Im $\alpha$ runs from $\gamma$ to $\infty$, and other with
 Re $\alpha=\theta'+\delta$ and Im $\alpha$ running from $\infty$ to $\gamma$, and a semi-circular path around
 $\alpha=\theta'+ i\gamma$. The latter vanishes in the limit $\delta\rightarrow 0$.
The presence of the square root in $R_{\alpha}$ introduces
a cut in the complex space that makes the contribution from the verticals paths the same.

The calculation of the second term is analogous to that made for the first one. Writing $\alpha=\theta'+i\beta$ to the up conribution and $\alpha=\theta'-i\beta$ to the bottom contribution, we obtain the same integral for both, so that last equation takes the form
\begin{eqnarray}
&&V(\rho,\theta,\varphi)=\frac{q}{8\pi^2\varepsilon_0}\frac{(\cosh\rho-\cos\theta)^{1/2}(\cosh\rho'-\cos\theta')^{1/2}}{a\sqrt{2}}\times \nonumber \\
&&\int\limits_{\gamma}^{\infty} (\cosh\beta-\cosh\gamma)^{-1/2}\sinh(\beta/2)\times\nonumber \\
&&\left[\cosh(\beta/2)-\cos\left(\frac{\theta-\theta'}{2}\right)\right]^{-1} d\beta \, .
\end{eqnarray}
We may perform the integration after defining the new variables
\begin{equation}
	\xi:=\cosh(\beta/2) \, , \sigma:=\cosh(\gamma/2) \, , \tau:=\cos\left(\frac{\theta-\theta'}{2}\right) \, .
	\label{xi}
\end{equation}
Doing that, we get
\begin{eqnarray}
	V(\rho,\theta,\varphi)&=&\frac{q}{8\pi^2\varepsilon_0}\frac{(\cosh\rho-\cos\theta)^{1/2}(\cosh\rho'-\cos\theta')^{1/2}}{a\sqrt{2}}\times\nonumber \\
&&\int\limits_{\sigma}^{\infty}\frac{d\xi}{(\xi^2-\sigma^2)^{1/2}(\xi-\tau)} \nonumber \\
\label{v2pronto}	&=& \frac{q}{2\pi^2\varepsilon_0R}\tan^{-1}\left[\frac{(\sigma+\tau)}{(\sigma-\tau)}\right]^{1/2} \, . \label{potumacarga}
\end{eqnarray}
Observe that at the point $\mathbf{r}=(\rho',\theta'+2\pi,\varphi')$, we have $\sigma=1$ and $\tau=1$, which makes
the potential $V$ divergent as $1/R$, while at $\mathbf{r}=(\rho',\theta'+2\pi,\varphi')$, we have $\sigma=1$ and
$\tau=-1$, which leaves the potential finite.

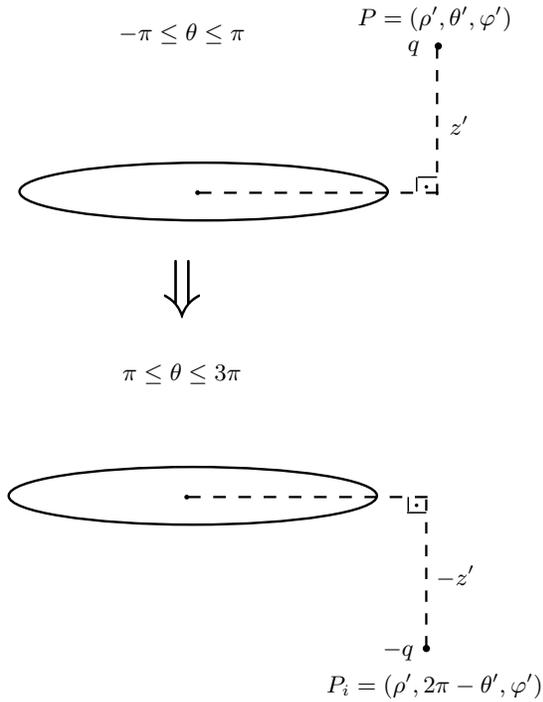
\begin{figure}
\begin{center}
\newpsobject{showgrid}{psgrid}{subgriddiv=1,griddots=10,gridlabels=6pt}
\begin{pspicture}(-1,-5.0)(9,5.94)
\psset{unit=0.8}
\psellipse[linewidth=0.04,dimen=outer](3.36,2.12)(3.08,0.5)
\usefont{T1}{ptm}{m}{n}
\rput(7.2,5){$P=(\rho',\theta',\varphi')$}
\psellipse[linewidth=0.04,dimen=outer](3.18,-2.94)(3.08,0.5)
\psdots[dotsize=0.12](7.26,4.54)
\usefont{T1}{ptm}{m}{n}
\rput(7.2,-6.1){$P_i=(\rho',2\pi-\theta',\varphi')$}
\psline[linewidth=0.03cm,linestyle=dashed,dash=0.16cm 0.16cm](7.24,4.52)(7.24,2.06)
\psdots[dotsize=0.12](7.06,-5.48)
\psdots[dotsize=0.08](3.26,2.1)
\psline[linewidth=0.03cm,linestyle=dashed,dash=0.16cm 0.16cm](7.06,-3.0)(7.06,-5.46)
\psline[linewidth=0.03cm,linestyle=dashed,dash=0.16cm 0.16cm](3.28,2.1)(7.26,2.1)
\psdots[dotsize=0.08](3.08,-2.96)
\psline[linewidth=0.02cm](6.9,2.12)(6.9,2.36)
\psline[linewidth=0.03cm,linestyle=dashed,dash=0.16cm 0.16cm](3.1,-2.96)(7.08,-2.96)
\psline[linewidth=0.02cm](6.92,2.36)(7.24,2.36)
\psline[linewidth=0.02cm](6.76,-3.2)(6.76,-2.96)
\psdots[dotsize=0.06](7.06,2.2)
\psline[linewidth=0.02cm](6.74,-3.22)(7.06,-3.22)
\usefont{T1}{ptm}{m}{n}
\rput(6.85,4.505){$q$}
\psdots[dotsize=0.06](6.9,-3.1)
\usefont{T1}{ptm}{m}{n}
\rput(3,4.745){$-\pi\leq\theta\leq\pi$}
\usefont{T1}{ptm}{m}{n}
\rput(6.59,-5.515){$-q$}
\usefont{T1}{ptm}{m}{n}
\rput(7.61,3.225){$z'$}
\usefont{T1}{ptm}{m}{n}
\rput(3,-0.915){$\pi\leq\theta\leq 3\pi$}
\usefont{T1}{ptm}{m}{n}
\rput(7.55,-4.275){$-z'$}
\rput(3,0.5){{\Huge $\Downarrow$}}
\usefont{T1}{ptm}{m}{n}
\rput(3.64,5.745){\color{white}aaaaaa}

\end{pspicture}
\end{center}
\caption{This figure shows both spaces, the real one and the imaginary auxiliary one. Note that while the real
charge $q$ is in the real space, the image charge $-q$ lies in the imaginary space.}
\label{disco}
\end{figure}

Potential $V$, given by equation (\ref{v2pronto}), satisfies Poisson equation but not the boundary conditions of the problem. Now we are entitled to find at which position of the double space we must put an image charge. We must put a charge $-q$ at the position
 $(\rho',2\pi-\theta',\varphi')$, as illustrate in Figure \ref{disco}. Then, the desired potential of the problem is given by the
 superposition of the potential of the real charge (located in the real space) with the potential of the image charge (located in the
 imaginary space), which turns to be
\begin{eqnarray}
		V_{disc}(\mathbf{r})&=&\frac{q}{2\pi^2\varepsilon_0}\Bigg\{R^{-1}\tan^{-1}\left[\frac{(\sigma+\tau)}{(\sigma-\tau)}\right]^{1/2}+ \nonumber \\
			&-& R^{-1}_{i}\tan^{-1}\left[\frac{(\sigma+\tau_i)}{(\sigma-\tau_i)}\right]^{1/2}\Bigg\} \, , \label{vpb}
\end{eqnarray}
with $R_i$ and $\tau_i$ obtained from $R$ and $\tau$ by the transformation $\theta\rightarrow 2\pi-\theta$,
\begin{eqnarray}
	\tau_i&=&-\cos\left(\frac{\theta+\theta'}{2}\right) \label{tauipbcap4} \\
R_i \!\!&=&  \!\! a\sqrt2\frac{[\cosh\gamma-\cos(\theta+\theta')]^{1/2}}{(\cosh\rho-\cos\theta)^{1/2}(\cosh\rho'-\cos\theta')^{1/2}} .
\end{eqnarray}
It is easy to see that in the conductor surface, $\theta=\pm\pi$, the boundary conditions, $V_{disc}=0$,
are obeyed.

The homogeneous Green function for this problem is given by equation (\ref{ghim}), with
\begin{eqnarray}
	\phi_i(\mathbf{r})&=&V_{disc}(\mathbf{r})-\frac{q}{4\pi |\mathbf{r}-\mathbf{r}'|}=\nonumber \\
	&=&-\frac{q}{2\varepsilon_0\pi^2}\Bigg\{R^{-1}\tan^{-1}\left[\frac{(\sigma-\tau)}{(\sigma+\tau)}\right]^{1/2}+\nonumber\\
		&+&R^{-1}_{i}\tan^{-1}\left[\frac{(\sigma+\tau_i)}{(\sigma-\tau_i)}\right]^{1/2}\Bigg\} \, ,
\end{eqnarray}
which leads to the result
\begin{eqnarray}
	G_H(\mathbf{r},\mathbf{r}')&=&-\frac{1}{2\pi^2}\Bigg\{R^{-1}\tan^{-1}\left[\frac{(\sigma-\tau)}{(\sigma+\tau)}\right]^{1/2}+ \nonumber \\
		&+& R^{-1}_{i}\tan^{-1}\left[\frac{(\sigma+\tau_i)}{(\sigma-\tau_i)}\right]^{1/2}\Bigg\} \, .
	\label{ghpb}
\end{eqnarray}

\begin{figure}[!h]
\centering
\includegraphics[scale=0.40]{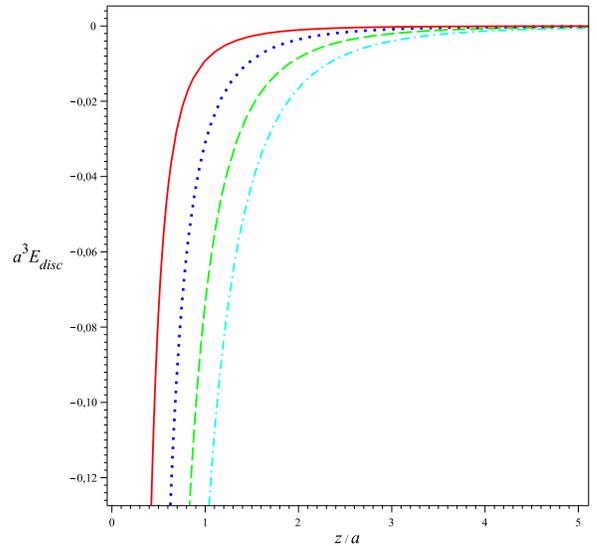}
\caption{Van der Waals interaction energy  for the atom-disc system (in a.u.), with the atom in the axis of the disc {\it versus}  $z/a$. Different curves correspond to different values of  $a$. For fixed $z$, greater values of $a$ yield more intense interactions. Note  the force on the atom is always attractive and it is a monotonic decreasing function of $z/a$.}
	\label{disco1}
\end{figure}

Substituting the previous results into equation (\ref{eberlein}) and taking the quantum expectation value, with the atom in its ground state, we get, after a lengthy but straightforward calculation, the non-retarded interaction energy between the atom and the disc. For an atom in the symmetry axis of the disc and polarizable predominantly in the direction of this axis, this interaction energy can be written in cylindrical coordinates in the form
\begin{eqnarray}
	E_{disc}=-\frac{\langle {\hat d}_z^2\rangle}{64\varepsilon_0\pi z^3}
\Biggl[\!\!\!\! &1&\! - \frac{2}{\pi}\sin^{-1}\left(\frac{z^2-a^2}{z^2+a^2}\right) + \cr
 &+& \!\frac{4az(3a^4+4a^2z^2+9z^4)}{3\pi(a^2+z^2)^3}\Biggr]\! . \label{adexato}
\end{eqnarray}
In (\ref{adexato}), $z$ is the atomic coordinate along the symmetry axis of the disc,
measured from its center, $a$ is the disc radius  and $\langle {\hat d}_z^2\rangle$ is the expectation value of the square of the (dominant) z-component of the  dipole operator $\hat{\bf d}$.
Figure \ref{disco1} shows the behavior of this interaction energy (conveniently multiplied by $a^3$ as a function of $z/a$.

>From  Eq.(\ref{adexato}), we can investigate the finite size effects for this system. Particularly, we can analyze how much the atom-disc system, with a finite radius $a$ for the disc, deviates from an atom interacting with an infinite plate for different distances between the atom and the center of the disc. Obviously, for very short distances, $z/a\ll 1$, the atom-disc system behaves like an atom-infinite plate system. However, as the ratio $z/a$ increases finite size effects start to become important. Whenever we have $z/a <1$, we can expand Eq.(\ref{adexato}) in powers of $z/a$ and consider more and more terms as $z/a$ becomes larger. The first terms of this expansion are given by
\begin{eqnarray} E_{disc} = E_0\Biggl[\!\!\!\! &1& - \frac{8}{3\pi}\left(\frac{z}{a}\right)^3\, + \cr
&+& \!\!\frac{48}{5\pi}\left(\frac{z}{a}\right)^5 - \frac{152}{7\pi}\left(\frac{z}{a}\right)^7 +
\mathcal{O}\left(\frac{z}{a}\right)^9\Biggr] \! , \label{adpert}
\end{eqnarray}
where $E_0=-\mbox{\large$\frac{\langle {\hat d}_z^2\rangle}{32\varepsilon_0\pi z^3}$}$ is the well known non-retarded dispersive interaction energy between the atom and a perfectly infinite conducting plate.

\begin{figure}[!h]
\centering
\includegraphics[scale=0.42]{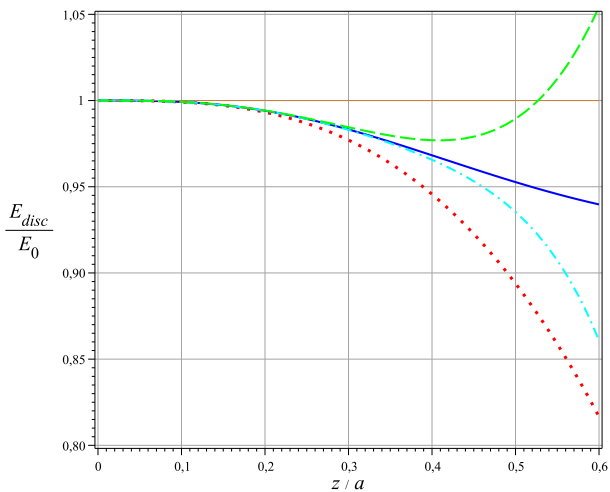}
\caption{Finite-size effects for an atom interacting with a disc. The solid line corresponds to the exact expression (\ref{adexato}), while the dotted, dashed and dotted-dashed curves correspond to the first three approximations, given by Eq.(\ref{adpert}).}
\label{disco3}
\end{figure}

In Figure \ref{disco3} we present the finite size effects for the atom-disc system by plotting the ratio $E_{disc}/E_0$ in several approximations, namely: continuous line shows the exact result (with $E_{disc}$ given by its exact expression (\ref{adexato})); dotted line shows the first approximation (with $E_{disc}$ given by the approximate expression (\ref{adpert}) up to the cubic term in $z/a$) and so on (dashed line corresponds to maintaining terms up to $(z/a)^5$ and dashed-dotted line, terms up to $(z/a)^7$).

\vskip -0.1 cm

\vskip -0.2cm
\noindent
A quick inspection in Figure \ref{disco3} (solid line) shows that for $z/a\approx 0.5$ the relative deviation of  $E_{disc}/E_0$ from  its unitary value when the disc is considered as an infinite plate  is of the order of $5\%$. Obviously, as the ratio $z/a$  increases this relative deviation becomes larger, since the finite size effects are more evident for larger values of $z/a$.

\section{Non-additivity in van der Waals interaction}

\begin{figure}[!h]
\centering
\includegraphics[scale=0.42]{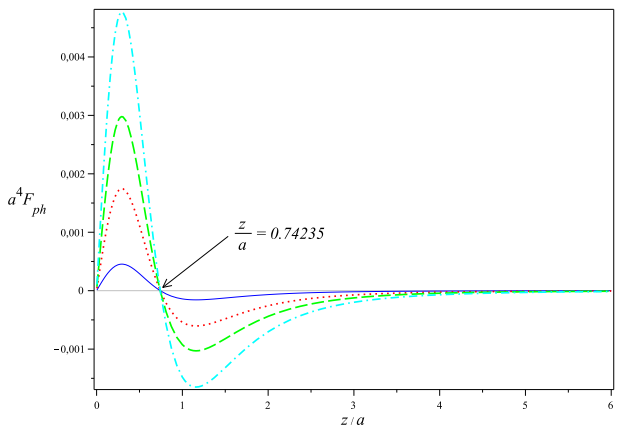}
\caption{Force on an atom polarizable only in the $z$ direction exerted by
	a perfectly conducting plate with a hole {\it versus} $z/a$ (in a.u.). The equilibrium position is indicated. For short distances we see repulsion.}
	\label{apb1}
\end{figure}

Non-additive effects are inherent to dispersive forces and have been known for a long time (see, for instance, Milonni's book \cite{Milonni-Book} and references therein). In a system of three atoms these effects were discussed in 1943 by Axilrod and Teller \cite{Axilrod-1943} and for $N$ molecules, in 1985, by Power and Thirunamachandran \cite{Power-1985}. Recently, non-additive effects were discussed for macroscopic bodies \cite{Rodriguez-PRA-2009,Ccapa-JPA-2010}. Here, our purpose is to discuss non-additivity of van der Waals interaction in systems involving an atom and complementary surfaces, like the above discussed atom-disc system and an atom interacting with an infinite conducting plate with a circular hole. With this goal, we shall also need the expression for the non-retarded interaction energy for the latter situation. This case was recently discussed by Eberlein and Zietal \cite{Eberlein2011} and its corresponding interaction energy can be obtained by the same method employed in the solution of the atom-disc system. Using Sommerfeld's image method and Eberlein-Zietal method, the non-retarded interaction energy between an atom and a perfectly conducting  plane with a circular hole can be written in the form
\vskip -0.5cm
\begin{eqnarray}
E_{ph} =-\frac{\langle {\hat d}_z^2\rangle}{64\varepsilon_0\pi z^3}\Biggl[\!\!\!\! &1& + \frac{2}{\pi}\sin^{-1}\left(\frac{z^2-a^2}{z^2+a^2}\right) \, +\cr\cr
&-& \frac{4az(3a^4+8a^2z^2-3z^4)}{3\pi(a^2+z^2)^3}\Biggr] \, .\label{apbexato}
\end{eqnarray}
In (\ref{apbexato}) all the symbols have the same meanings as in the atom-disc case, except for
$a$, which now stands for the radius of the circular hole. (Eq.(\ref{apbexato}) is equivalent to Eberlein and Zietal's result).
%
%
In Figure \ref{apb1} we plot the non-retarded dispersive force on the atom exerted by the infinite plate with a circular hole. Different curves correspond to different values of $a$. Note that the position of the atom which leads to stable equilibrium (in the symmetry axis) depends only on the ratio $z/a$ and it occurs for $z\approx 0,74235a$ as already pointed oud in \cite{Eberlein2011}. It is also proved that this equilibrium is unstable under lateral displacements, as expected.
%

The results exposed previously render a proper picture to the study of non-additivity in the van der Waals forces. With this goal, observe initially that the derivative of equation (\ref{adexato}) with respect to $z$  gives us the van der Waals force on the atom exerted by the disc,
 $F_{disc} = -\partial_z E_{disc}$, while the derivative of equation (\ref{apbexato}) with respect to $z$  gives us the van der Waals force on the atom exerted by the conducting plane with a circular hole,  $F_{ph} = -\partial_z E_{ph}$. To estimate the non-additive effects means to quantify how much the superposition $F_{disc} + F_{ph}$ differs from the van der Waals force on the atom exerted by an infinite conducting plane (superposition of the two complementary surfaces), given by $F_0 = -\partial_z E_0$. From equations (\ref{adexato}) and (\ref{apbexato}) is straightforward to show that
\begin{equation}
F_{disc}+F_{ph} = F_0 - \dfrac{\langle d_z^2\rangle a}{\varepsilon_0\pi^2} \dfrac{(z^2-a^2)z}{(z^2+a^2)^4}
\label{NAforce}
\end{equation}

\begin{figure}[!h]
\centering
\includegraphics[scale=0.36]{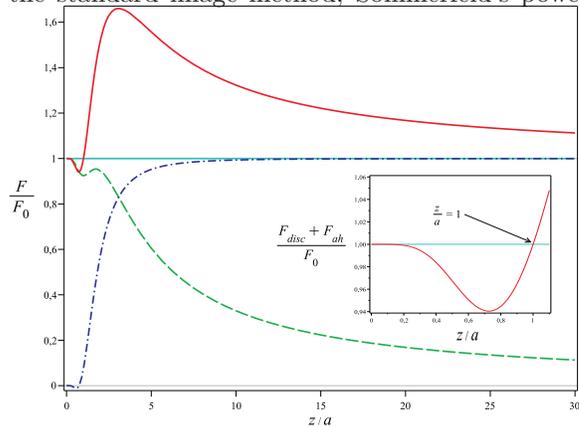}
\caption{Non-additivity effects in complementary systems. The dashed curve corresponds to $F_{disc}/F_0$ {\it versus} $z/a$, while the
dotted-dashed curve stands for $F_{ph}/F_0$ {\it versus} $z/a$. The solid curve is the sum of these two curves.}
\label{disco4}
\end{figure}

The last term in the rhs of (\ref{NAforce}) corresponds to the non-additivity term. In Figure \ref{disco4} the solid line shows the behavior of this non-additivity term (divided by $F_0$) as a function of the ratio $z/a$. As expected, for $z/a \rightarrow 0$  and  $z/a \rightarrow\infty$ the non-additivity term disappears. Naively, one could expect that the non-additivity effects started from zero (for $z/a\rightarrow 0$), increased and then decreased to zero (for $z/a\rightarrow\infty$). Curious as it may seem, a quite unexpected result occurs, namely,  the non-additivity term vanishes at $z=a$ or, equivalently, $(F_{disc} + F_{ph})/F_0 = 1$ for $z=a$ (see the detail in the box of Figure \ref{disco4}). Though this result was obtained analytically from the previous equations, there is a qualitative argument to understand it. It is not difficult to show that for $z/a\ll 1$, $(F_{disc} + F_{ph})/F_0$ is slightly smaller than one, while for $z/a\gg 1$, it is slightly greater than one, so that it will necessarily assune the unitary value for a finite value of $z/a$ (once we assume $(F_{disc} + F_{ph})/F_0$ is a continuous function for $0<z/a<\infty$). It can also be shown that these arguments seem to be independent of the form of the hole, so that we are tempted to conjecture that, for any hole possessing a symmetry axis, such as any regular polygon, there will exist a point on this axis at which the non-additivity associated to complementary surfaces  disappears.


%

%
%
\section{Final remarks}

In this paper we presented the exact result for the van der Waals interaction energy between an atom and a perfectly conducting disc. For this purpose, we have used a method recently developed by Eberlein and Zietal \cite{Eberlein2007} combined with the image method, which yields a rather simple approach to deal with atoms near conductors. Although the disc configuration is not treatable by the standard image method,
Sommerfeld's powerful extension of this method allowed us to treat it. In Sommerfeld's approach, the image charge is located in a copy of the ordinary space. We discussed the finite size effects for such a non-trivial system and used our results, together with the recently results obtained in \cite{Eberlein2011} for the van der Waals interaction energy between an atom and a perfectly conducting infinite plate with a circular hole to discuss non-additivity effects in the van der Waals interaction involving complementary surfaces. We found a very peculiar result, namely, that there exists a given ratio $z/a$ for which the non-additivity effect completely disappears. There are qualitative arguments which suggest that this quite unexpected result may occur to other pairs of planar complementary geometries.

We would like to emphasize that Sommerfeld's image method is very well suited for calculations of van der Waals interactions by using Eberlein-Zietal method. Here we used it in two situations, but it can be used in other situations, provided the problem in consideration admits a solution via image method. As a final comment, we believe that the atom-disc system and the atom interacting with the complementary surface (an infinite plate with a circular hole) should be used in further investigations under the light of Babinet's Principle in dispersive interactions, in the spirit of the discussion presented by Maghrebi et al \cite{Maghrebi2011}.

\noindent
{\bf Acknowledgements}\\
The authors are indebted with P.A. Maia Neto, F.S.S. Rosa, F. Pinheiro and A.L.C. Rego for
valuable discussions. The authors  also  thank to CNPq and FAPERJ (brazilian agencies) for
partial financial support.

%
%

%
\end{document}